\documentstyle[
aps]{revtex}
\def\be{\begin{equation}}
\def\ee{\end{equation}}
\def\bea{\begin{eqnarray}}
\def\eea{\end{eqnarray}}
\def\ba{\begin{array}}
\def\ea{\end{array}}
\def\nn{\nonumber}
\def\p{\partial}
\def\ka{\kappa}
\def\ep{\epsilon}
\def\sta{\sin\theta}

\def\sda{\sin^2\theta}
\def\cda{\cos^2\theta}

\begin{document}
\pagestyle{myheadings}
\markboth{}{Wu and Cai}
\draft
\twocolumn[\hsize\textwidth\columnwidth\hsize\csname 
           @twocolumnfalse\endcsname

\title{Massive Complex Scalar Field in a Kerr-Sen Black Hole Background:\\
Exact Solution of Wave Equation and Hawking Radiation}
\author{S. Q. Wu$^{1)}$ and X. Cai$^{2)}$}
\address{Institute of Particle Physics, Hua-Zhong 
Normal University, Wuhan 430079, P. R. China \\ 
\rm $^{1)}$E-mail: sqwu@iopp.ccnu.edu.cn ~~~~ 
$^{2)}$E-mail: xcai@ccnu.edu.cn}
\date{\today}
\maketitle

\begin{abstract}

The separated radial part of a massive complex scalar wave equation in 
the Kerr-Sen geometry is shown to satisfy the generalized spheroidal wave 
equation which is, in fact, a confluent Heun equation up to a multiplier. 
The Hawking evaporation of scalar particles in the Kerr-Sen black hole 
background is investigated by the Damour-Ruffini-Sannan's method. It is
shown that quantum thermal effect of the Kerr-Sen black hole has the same
characteras that of the Kerr-Newman black hole.

 
\end{abstract}
]

\narrowtext

\section{Introduction}

Exact solutions to various wave equations such as Klein-Gordon, Dirac and 
Maxwell equations in a special black hole background are of considerable 
importance in mathematical physics as well as in black hole physics. 
Knowing an explicit expression for some exact solutions is undoubtedly 
very helpful to construct Feynman Green function inside a Schwarzschild 
black hole \cite{CJ}, to do second quantization in the Kerr metric 
\cite{Guven}, to study quasi-normal modes of Kerr black hole and 
Kerr-Newman black hole \cite{LOK}, to check stability of Kerr black 
hole and Kerr-de Sitter spacetime \cite{WCM}, to calculate absorption 
rate of Kerr-de Sitter black hole and Kerr-Newman-de Sitter black hole 
\cite{STU1} and to investigate black hole perturbation \cite{MST}, etc. 
Effort to seek such exact solutions has been undertaken all the time. 
It is instructive to present a brief review on studies about exact 
solution of a quantum field equation in some well-known black hole 
spacetimes.

At 1982, Birrell and Davies \cite{BD} stated in their classical masterpiece 
{\em Quantum Fields in Curved Space} that \lq\lq It is not possible to 
write the solutions $R_{\omega\ell}(r)$ to the radial equation (8.2) in 
terms of known functions, though the properties of the solutions have 
been extensively investigated." (see also \cite{PLP}). By the radial 
equation, they referred to the separated radial part of a massless 
Klein-Gordon equation in the Schwarzschild space-time. Prior to that 
time, exact solution to the angular part of a wave equation on a black 
hole background had already received a large mount of studies and named 
as the spin-weighted angular spheroidal wave function \cite{AWE}, but 
solution to its radial part \cite{RWE} was unknown to almost all of 
those researchers who studied black hole physics. The radial wave 
equation was thought of as not being related to any known differential 
equation of mathematical physics before then \cite{PLP}.

This situation changed in the next year. Blandin, Pons and Marcilhacy 
\cite{BPM} showed that the spin-weighted spheroidal functions may be 
obtained by an elementary transformation from Heun's confluent function. 
They also noted that both the angular part and the radial part of 
Teukolsky's master equation \cite{TPT} are particular forms of a single 
linear ordinary differential equation. By a private communication, G. W. 
Gibbons notified Jensen and Candelas \cite{JC} that the radial part for 
a massless scalar wave equation in the Schwarzschild spacetime is, up to 
a multiplier, a confluent Heun function.

The next important development started from Leaver's work \cite{Leaver}. 
He showed that the radial and angular parts of Teukolsky master equation
\cite{TPT} are equations of the same type, namely, the generalized spheroidal 
wave equation (GSWE) \cite{Leaver,LFN}. Especially, he also expanded the
solution of a GSWE in terms of hypergeometry and confluent hypergeometry 
functions. Recently, we demonstrate that both the radial part and the angular 
part of a massive charged scalar field equation on the Kerr-Newman black 
hole background can be transformed into a GSWE \cite{WC}.

The most remarkable progress comes from recent researches on revealing the 
relations among the generalized Teukolsky master equation, GSWE and Heun 
equation. Suzuki, Takasugi and Umetsu \cite{STU2} have studied perturbations 
of Kerr-de Sitter black holes and analytic solutions of Teukolsky equation 
in the Kerr-de Sitter and Kerr-Newman-de Sitter geometries, and shown that 
this generalized Teukolsky master equation has a close relation to Heun's 
equation \cite{Ron} (and references therein) which is introduced by Heun 
\cite{Heun} as early as 1889. Also, they have proved that a GSWE is a confluent 
Heun equation when cosmological constant approaches to zero. Analytic solutions 
of the Teukolsky equation were studied in Ref. \cite{AWE,STM}. The integral 
equations of fields on the rotating black hole had been investigated in 
Ref. \cite{MS}. The integral equations for Heun functions were presented 
in Ref. \cite{LWEV}.

Through these researches, it may be generally accepted that the generalized 
Teukolsky master equation in the Kerr-Newman-de Sitter spacetimes can be 
recast into the form of a generalized spheroidal wave equation which is, in 
fact, a Heun equation. But it is not clear until now whether the generalized 
Teukolsky equation in the general type-D vaccum backgrounds with cosmological 
constant (namely, stationary C-metric solutions \cite{PD}) can be transformed 
into a Heun equation. Here we also mention Couch's work \cite{Couch} that 
relates a series of exact solutions by transforming the separated radial 
equation into a modified Whittaker-Hill equation under some special conditions. 

Ina recent paper \cite{WC}, we have investigated exact solution of a massive 
complex scalar field equation in the Kerr-Newman black hole background. 
Previous work on solution of a massive scalar wave equation in the Kerr(-Newman) 
spacetime had been completed in Ref. \cite{KGK}. It is interesting to extend our analysis 
to solution of a scalar wave equation in a Kerr-Sen black hole background 
\cite{KS}. The Kerr-Sen solution \cite{KS} arising in the low energy effective 
string field theory is a rotating charged black hole generated from the Kerr 
solution. The thermodynamic property of this twisted Kerr black hole was 
discussed in Ref. \cite{Okai} by using separation of the Hamilton-Jacobi 
equation of a test particle. The aim of this paper is to find some exact 
solutions to a massive charged scalar wave equation and to investigate 
quantum thermal effect of scalar particles on the Kerr-Sen spacetime.
 
The paper is organized as follows: In Sec. 2, we separate a massive charged 
scalar field equation on the Kerr-Sen black hole background into the radial 
and angular parts. Sec. 3 is devoted to transforming the radial part into a 
generalized spheroidal wave equation and to relating it to the confluent Heun 
equation. Then, we investigate quantum thermal effect of scalar particles in 
the Kerr-Sen spacetime in Sec. 4. Finally, we summarize our discussions in 
the conclusion section.

\section{Separating variables of Klein-Gordon 
equation on Kerr-Sen black hole background}

Constructed from the charge neutral rotating (Kerr) black hole solution, 
the Kerr-Sen solution \cite{KS} is an exact classical four dimensional 
black hole solution in the low energy effective heterotic string field 
theory. In the Boyer-Lindquist coordinates, the Kerr-Sen metric can be 
rewritten as \cite{Okai}:
\bea
ds^2 &=& -\frac{\Delta}{\Sigma}\Big(dt -a\sda 
d\varphi\Big)^2 +\frac{\sda}{\Sigma}\Big[adt \nn \\
&&-(\Sigma +a^2\sda)d\varphi\Big]^2 +\Sigma
\Big(\frac{dr^2}{\Delta} +d\theta^2\Big) \, ,
\eea
where $\Delta = r^2 +2(b -M)r +a^2 = (r -r_+)(r -r_-)$, $\Sigma = r^2 +2br 
+a^2\cda$ and $r_{\pm} = M -b \pm \ep$ with $\ep = \sqrt{(M -b)^2 -a^2}$. 
The electromagnetic field vector potential is concisely chosen as \cite{Okai}
\be
{\cal A} = \frac{-Qr}{\Sigma}(dt -a\sda d\varphi) \, .
\ee

This metric describes a black hole carrying mass $M$, charge $Q$, angular 
momentum $J = Ma$, and magnetic dipole moment $Qa$. The twist parameter
$b$ is related to the Sen's parameter $\alpha$ via $b = Q^2/2M = M \tanh^2
(\alpha/2)$. Because $M \geq b \geq 0$, $r = r_-$ is a new singularity in 
the region $r \leq 0$, the event horizon of the Kerr-Sen black hole is 
located at $r = r_+$. The area of the out event horizon of the twisted 
Kerr solution \cite{KS} is given by ${\cal A} = 4\pi(r_+^2 +2br_+ +a^2) 
= 8\pi Mr_+$.

We consider the solution of a massive charged test scalar field on the
Kerr-Sen black hole background (In Planck unit system $G = \hbar = c = 
k_B = 1$). The complex scalar field $\Phi$ with mass $\mu$ and charge $q$ 
in such a space-time satisfies the following Klein-Gordon equation (KGE):
\bea
&&\frac{-1}{\Delta}\Big[(r^2 +2br +a^2)\p_t +a\p_{\varphi} 
+iqQr\Big]^2\Phi  \nn \\
&&~~~~ +\p_r(\Delta\p_r\Phi)+\frac{1}{\sta}\p_{\theta}(\sta\p_{\theta}\Phi) 
\nn \\
&&~~~~ +\Big(a\sta\p_t +\frac{1}{\sta}\p_{\varphi}\Big)^2\Phi 
= {\mu}^2\Sigma\Phi \, . 
\eea

The wave function $\Phi$ of KGE has a solution of variables separable form 
$\Phi(t,r,\theta,\varphi) = R(r)S(\theta)e^{i(m\varphi -\omega t)}$, in which 
the separated radial and angular parts of KGE can be given as follows:
\bea
&&\frac{1}{\sta}\p_{\theta}[\sta\p_{\theta}S(\theta)]
+\Big[\lambda -\frac{m^2}{\sda} \nn \\
&&~~~~~~~~~~~~~~~ +({\mu}^2 -{\omega}^2)a^2 \sda\Big]S(\theta) = 0 \, ,
\label{ane} \\
&&\p_r[\Delta\p_rR(r)] +\Big[\frac{K^2(r)}{\Delta} -{\mu}^2(r^2 
+2br +a^2) \nn \\
&&~~~~~~~~~~~~~~~~~~~~~~~~~ -\lambda +2ma\omega\Big]R(r) = 0 \, , 
\label{rae}
\eea
here $\lambda$ is a separation constant, $K(r) = \omega(r^2 +2br +a^2)
-qQr -ma$.    

The general solution $S_{m,0}^{\ell}(ka,\theta)$ to the angular equation 
(\ref{ane}) is an ordinary spheroidal angular wave functions \cite{OSWF} 
with spin-weight $s = 0$, while the radial equation (\ref{rae}) can be 
reduced to 
\bea
&&\p_r[\Delta\p_rR(r)] +\Big[\frac{(Ar -ma)^2}{\Delta} +k^2\Delta \nn \\
&&~~~~~~~~~~~~~~~~~~~~~ +2Dr -\lambda \Big]R(r) = 0 \, , \label{rew}
\eea
where we have put $A = 2M\omega -qQ$, $D = A\omega -M{\mu}^2$, $k =
\sqrt{{\omega}^2 -{\mu}^2}$ (supposed that $\omega > \mu$). For later 
convenience, we also denote $\ep B = A(M-b) -ma$ and introduce $W_{\pm} 
= (A \pm B)/2$. 

Eq. (\ref{rew}) has two regular singular points $r = r_{\pm}$ with indices 
$\pm iW_+$ and $\pm iW_-$, respectively. The radial wave function $R(r)$ 
has asymptotic behaviors when $r \rightarrow r_{\pm}$: 
\be 
R(r) \sim \left\{
\ba{l}
(r -r_+)^{\pm iW_+} \, , ~~~~ (r \rightarrow r_+)  \\
(r -r_-)^{\pm iW_-} \, , ~~~~ (r \rightarrow r_-) 
\ea \right.
\ee

Making substitution $R(r) = (r -r_+)^{i(A +B)/2}(r -r_-)^{i(A -B)/2}F(r)$, 
we can transform Eq. (\ref{rew}) for $R(r)$ into a modified generalized 
spheroidal wave equation with imaginary spin-weight $iA$ and boost-weight
$iB$ for $F(r)$ \cite{WC}:
\bea
&&\Delta\p_r^2F(r) +2[i\epsilon B +(1 +iA)(r -M +b)]\p_rF(r) \nn \\
&&~~~~~~~~~~~~~~~~~~~~ +[k^2\Delta +2Dr +iA -\lambda]F(r) = 0 \, . 
\label{gswe}
\eea

Eq. (\ref{gswe}) has indices $\rho_+ = 0$, $-2iW_+$ and $\rho_- = 0$, $-2iW_-$
at two singularities $r = r_{\pm}$, respectively. Thus there have two linear
independent solutions at each point. The infinity is an irregular singularity 
of Eq. (\ref{rew}) or (\ref{gswe}). The functions $R(r)$ and $F(r)$ at infinity 
have asymptotic form $R(r)\sim e^{\pm ikr}$. Eq. (\ref{gswe}) has the same
form as the radial part of the massive complex scalar wave equation in the
Kerr-Newman geometry \cite{WC} with its solution \cite{Leaver} (when $\mu = 0$)  
named as the generalized spheroidal wave function. It is interesting to note 
that a special solution of function $F(r)$ satisfies the Jacobi equation of 
imaginary index when $\omega = \pm \mu = qQ/M$ (namely, $k =D = 0$).

\section{Generalized spheroidal 
wave function and Heun equation}

In this section, we shall show that the generalized spheroidal wave equation 
(\ref{gswe}) of imaginary number order is, in fact, a confluent form of Heun 
equation \cite{Ron}. To this end, let us make a coordinate transformation 
$r = M -b +\ep z$ and substitute $R(r) = (z -1)^{i(A +B)/2}(z +1)^{i(A -B)/2}
F(z)$ into Eq. (\ref{rew}), then we can reduce it to the following standard 
forms of a generalized spheroidal wave equation \cite{WC,Ron}:
\bea
&&(z^2 -1)R^{\prime\prime}(z) +2zR^{\prime}(z) +\Big[(\ep k)^2(z^2 -1)
+2D\ep z \nn \\
&&~~~~~~~~ +\frac{(Az +B)^2}{z^2 -1} +2D(M -b) -\lambda\Big]R(z) = 0 \, ,
\eea
and
\bea
&&(z^2 -1)F^{\prime\prime}(z) +2[iB +(1 +iA)z]F^{\prime}(z) 
+[(\ep k)^2(z^2 -1) \nn  \\
&&~~~~~~~~~~~ +2D\ep z +2D(M -b) +iA -\lambda]F(z) = 0 \, , 
\label{swe} 
\eea
where a prime denote the derivative with respect to its argument.

The spin-weighted spheroidal wave function $F(z)$ is symmetric under the 
reflect $k \rightarrow -k$. Letting $F(z) = e^{i\ep kz}G(z)$ without loss 
of generality, we can transform Eq. (\ref{swe}) to
\bea
&&(z^2 -1)G^{\prime\prime}(z) +2[iB +(1 +iA)z 
+i\ep k(z^2 -1)]G^{\prime}(z) \nn \\
&&~~~~~~~~~ +[2i\ep k(1 +iA -iD/k)z -2\ep kB +iA  \nn \\
&&~~~~~~~~~~~~~~~~~~~~~~ +2D(M -b) -\lambda]G(z) = 0 \, . \label{tswe}
\eea
By means of changing variable $z = 1 -2x$, we arrange the singularities
$r = r_+$ ($z = 1$) to $x = 0$ and $r = r_-$ ($z = -1$) to $x = 1$,
respectively, and reduce Eq. (\ref{tswe}) to a confluent form of Heun's  
equation \cite{BPM,Ron} 
\be
G^{\prime\prime}(x) +\Big(\beta +\frac{\gamma}{x} +\frac{\delta}{x-1} 
\Big)G^{\prime}(x) +\frac{\alpha\beta x -h}{x(x -1)}G(x) = 0 \, ,
\label{cH}
\ee
with $\gamma = 1 +2iW_+$, $\delta = 1 +2iW_-$, $\beta = 4i\ep k$, 
$\alpha = -(1 +iA) +iD/k$, $h = \lambda -2i\ep k -iA +4\ep kW_+ -2Dr_+$.

This confluent Heun equation (\ref{cH}), with $h$ its accessory parameter,
has two regular singular points at $x = 0, 1$ with exponents ($0, 1 -\gamma$)
and ($0, 1 -\delta$),respectively, as well as an irregular singularity at 
the infinity point. The power series solution in the vicinity of the point 
$x = 0$ for Eq. (\ref{cH}) can be written as
\be
G(\alpha,\beta,\gamma,\delta,h;x) = \sum\limits_{n = 0}^{\infty}g_n x^n \, ,
\ee
and the coefficient $g_n$ satisfies a three-term recurrence relation
\cite{BPM,Ron}
\bea
&&g_0 = 1 \, , ~~~~ g_1 = -h/\gamma \, , \nn \\
&&(n +1)(n +\gamma)g_{n +1} -\beta(n -1+\alpha)g_{n -1} \nn \\
&&~~~~~~~~  = \big[n(n -1 -\beta +\gamma +\delta) -h\big]g_n  \, .
\eea
It is not difficult to deduce the exponent $1 -\gamma$ solution for $x = 0$ 
\cite{BPM} and obtain the power series solution in the vicinity of the point 
$x = 1$ by a linear transformation interchanging the regular singular points 
$x = 0$ and $x = 1$: $x \rightarrow 1 -x$. Expansion of solutions to the 
confluent Heun's equation in terms of hypergeometric and confluent 
hypergeometric functions has been presented in \cite{Ron,STM,Leaver}. The 
confluent Heun's functions can be orthonormalized to constitute a group of 
orthogonal complete functions \cite{Ron}. It should be noted that Heun's 
confluent equation also admits quasi-polynomial solutions for particular 
values of the parameters \cite{BPM,Ron}. It follows from the three-term 
recurrence relation that $G(\alpha,\beta,\gamma,\delta,h;x)$ is a polynomial 
solution if 
\bea
&&\alpha = -N \, , ~~~~{\rm with~~ integer}~~ N \geq 0 \, , \nn \\
&&g_{N +1}(h) = 0 \, , 
\eea
$g_{N +1}$ being a polynomial of degree $N +1$ in $h$, that is, therefore  
$N +1$ eigenvalues $h_i$ for $h$ such as $g_{N +1}(h_i) \equiv 0$.

\section{Hawking radiation of scalar particles}

Now we investigate the Hawking evaporation \cite{Hawk} of scalar particles
in the Kerr-Sen black hole by using the Damour-Ruffini-Sannan's (DRS) method 
\cite{DRS}. This approach only requires the existence of a future horizon and 
is completely independent of any dynamical details of the process leading to 
the formation of this horizon. The DRS method assumes analyticity properties 
of the wave function in the complexified manifold.

In the following, we shall consider a wave outgoing from the event horizon 
$r_+$ over interval $r_+ <r <\infty$. According to the DRS method, a correct 
outgoing wave $\Phi^{\rm out} = \Phi^{\rm out}(t,r,\theta,\varphi)$ is an 
adequate superposition of functions $\Phi_{r >r_+}^{\rm out}$ and $\Phi_{r 
<r_+}^{\rm out}$:
\be
\Phi^{\rm out} = C\big[\eta(r -r_+)\Phi_{r >r_+}^{\rm out}
+\eta(r_+ -r)\Phi_{r <r_+}^{\rm out}e^{2\pi W_+}\big] \, ,
\ee
where $\eta$ is the conventional unit step function, $C$ is a 
normalization factor.

In fact, components $\Phi_{r >r_+}^{\rm out}$ and $\Phi_{r <r_+}^{\rm out}$
have asymptotic behaviors: 
\bea
\Phi_{r >r_+}^{\rm out} &=& \Phi_{r >r_+}^{\rm out}(t,r,\theta,\varphi) 
\longrightarrow c_1(r -r_+)^{iW_+} \nn \\
&& \times S_{m,0}^{\ell}(ka,\theta)e^{i(m\varphi -\omega t)} \, ,
~~~~ (r \rightarrow r_+) \\ 
\Phi_{r <r_+}^{\rm out} &=& \Phi_{r <r_+}^{\rm out}(t,r,\theta,\varphi) 
\longrightarrow c_2(r -r_+)^{-iW_+} \nn \\
&& \times S_{m,0}^{\ell}(ka,\theta) e^{i(m\varphi -\omega t)} \, ,
~~~~ (r \rightarrow r_+)
\eea
when $r \rightarrow r_+$. Clearly, the outgoing wave $\Phi_{r >r_+}^{\rm 
out}$ can't be directly extended from $r_+ <r <\infty$ to $r_- <r <r_+$, 
but it can be analytically continued to an outgoing wave $\Phi_{r <r_+}^{\rm 
out}$ that inside event horizon $r_+$ by the lower half complex $r$-plane 
around unit circle $r = r_+ -i0$:
$$ r -r_+ \longrightarrow (r_+ -r)e^{-i\pi} \, .$$
By this analytical treatment, we have
\be
\Phi_{r <r_+}^{\rm out}\sim c_2(r -r_+)^{-iW_+} 
S_{m,0}^{\ell}(ka,\theta)e^{i(m\varphi -\omega t)} \, .
\ee

As a difference factor $(r -r_+)^{-2iW_+}$ emerges, then $\Phi_{r >r_+}^{\rm 
out}$ differs $\Phi_{r <r_+}^{\rm out}$ by a factor $e^{2\pi W_+}$, thus we 
can derive the relative scattering probability of the scalar wave at the
event horizon
\be
\Big|\frac{\Phi_{r >r_+}^{\rm out}}{\Phi_{r <r_+}^{\rm out}}\Big|^2 
= e^{-4\pi W_+} \, ,
\ee
and obtain the thermal radiation spectrum with the Hawking temperature 
$T = \ka/2\pi$. 
\bea
&&\langle {\cal N} \rangle = |C|^2 = \frac{1}{e^{4\pi W_+} -1} \, ,
\label{sptr} \\
&&W_+ = \frac{Ar_+ -ma}{2\ep}= \frac{\omega -m\Omega -q\Phi}{2\ka} \, ,
\eea
where the angular velocity at the horizon is $\Omega = a/2Mr_+$, the 
electric potential is $\Phi = Q/2M = b/Q$, the surface gravity at the 
pole is $\ka = (r_+ -M +b)/2Mr_+ = \ep/2Mr_+$. 

The black body radiation spectrum (\ref{sptr}) demonstrates that the thermal 
property of Kerr-Sen black hole is similar to that of Kerr-Newman black hole 
though its geometry character likes that of the Kerr solution \cite{Okai}. 
Correspondingly, there exist four thermodynamical laws of the Kerr-Sen black 
hole, similar to those of Kerr-Newman black hole thermodynamics. 

\section{Conclusion}

In this paper, we have shown that the separation of variables of the scalar
wave equation in the Kerr-Newman black hole background \cite{WC,KGK} can
apply completely to the case of the twisted Kerr solution. The separated 
radial part can be recast into the generalized spheroidal wave equation, 
which is, in fact, a confluent form of Heun equation.
   
In addition, we find that the thermal property of the twisted Kerr black hole 
resembles that of Kerr-Newman black hole though its geometry character likes 
that of the Kerr solution. The Kerr-Sen solution shares similar four black hole 
thermodynamical laws and quantum thermal effect as the Kerr-Newman spacetime
does.

One of us (SQW) is very indebted to Dr. Jeff Zhao and Dr. C. B. Yang for
their helps in finding some useful references. He also thanks Prof. E. 
Takasugi and Prof. G. Valent very much for kindly sending their reprints.
This work is supported partially by the NSFC in China under Grant number: 
19875019.

  
  

\begin{thebibliography}{s2}  

\bibitem{CJ}
P. Candelas and B. P. Jensen, {\it Phys. Rev}. {\bf D33} 
(1986) 1596.

\bibitem{Guven}
R. G\"uven, {\it Phys. Rev}. {\bf D10} (1974) 3194.

\bibitem{LOK}
E. W. Leaver, {\it Proc. R. Soc. Lond}. {\bf A402} (1985) 285; 
H. Onozawa, {\it Phys. Rev}. {\bf D55} (1997) 3593; 
K. D. Kokkotas, {\it IL Nuovo Cimento}, {\bf B108} (1993) 991. 

\bibitem{WCM}
B. F. Whiting, {\it J. Math. Phys}. {\bf 30} (1989) 1301; 
C. M. Chambers and I. G. Moss, {\it Class. Quant. Grav}. 
{\bf 11} (1994) 1035.

\bibitem{STU1}
H. Suzuki, E. Takasugi and H. Umetsu, {\it Prog. Theor. Phys}. 
{\bf 103} (2000) 723.

\bibitem{MST}
Y. Mino, M. Sasaki, M. Shibata, H. Tagoshi and T. Tanaka, 
{\it Prog. Theor. Phys. Suppl}. {\bf 128} (1997) 1.

\bibitem{BD} 
N. D. Birrell and P. C. W. Davies, {\it Quantum Fields in 
Curved Space}, (Cambridge University Press, Cambridge, 1982).

\bibitem{PLP}
S. Persides, {\it J. Math. Phys}. {\bf 14} (1973) 1017; 
D. Lohiya and N. Panchapakesan, {\it J. Phys. A: Math. Gen}. 
11 (1978) 1963.

\bibitem{AWE}
W. H. Press and S. A. Teukolsky, {\it Astrophys. J}. 
{\bf 185} (1973) 649; 
R. A. Breuer, M. P. Ryan, S. Waller, {\it Proc. R. Soc. Lond}. 
{\bf A358} (1977) 71; 
E. D. Fackerell, R. G. Crossman, {\it J. Math. Phys}. {\bf 18} 
(1977) 1849; 
K. G. Suffern, E. D. Fackerell and C. M. Cosgrove, {\it J. Math. 
Phys}. {\bf 24} (1983) 1350; 
D. A. Leahy and W. G. Unruh, {\it Phys. Rev}. {\bf D19} (1979) 
3509; 
S. K. Chakrabarti, {\it Proc. R. Soc. Lond}. {\bf A391} (1984) 
27; 
E. Seidel, {\it Class. Quant. Grav}. {\bf 6} (1989) 1057; 
E. G. Kalnins, W. Jr. Miller, {\it J. Math. Phys}. 
{\bf 33} (1992) 286; 
B. P. Jensen, J. G. Mc Laughlin and A. C. Ottewill, 
{\it Phys. Rev}. {\bf D51} (1995) 5676.

\bibitem{RWE}
J. M. Bardeen and W. H. Press, {\it J. Math. Phys}. 
{\bf 14} (1973) 7; 
D. Page, {\it Phys. Rev}. {\bf D13} (1973) 198; 
C. H. Lee, {\it J. Math. Phys}. {\bf 17} (1976) 1226; 
{\it Phys. Lett}. {\bf 68B} (1977) 152; 
R. F. Arenstorf, J. M. Cohen and L. S. Kegeles, 
{\it J. Math. Phys}. {\bf 19} (1978) 833; 
B. R. Iyer and A. Kumar, {\it Pramana}, {\bf 11} (1978) 171; 
R. G\"uven, {\it Phys. Rev}. {\bf D16}. (1977) 1706; 
A. Zecca, {\it IL Nuovo Cimento}, {\bf B113} (1998) 1309; 
{\bf B115} (2000) 625; 
B. Mukhopadhyay and S. K. Chakrabarti, {\it Nucl. Phys}. 
{\bf B582} (2000) 627; 
S. K. Chakrabarti and B. Mukhopadhyay, {\it IL Nuovo Cimento}, 
{\bf B115} (2000) 885; astro-ph/0007277. 

\bibitem{BPM}
J. Blandin, R. Pons and G. Marcilhacy, {\it Lett. Nuovo Cimento}, 
{\bf 38} (1983) 561.

\bibitem{TPT}
S. A. Teukolsky, {\it Astrophys. J}. {\bf 185} (1973) 635; 
{\it Phys. Rev. Lett}. {\bf 29} (1972) 1114; 
S. A. Teukolsky and W. H. Press, {\it Astrophys. J}. 
{\bf 193} (1974) 443; 
S. Chandrasekhar, {\it The Mathematical Theory of Black Holes}, 
(New York: Oxford University Press, 1983); 
U. Khanal, {\it Phys. Rev}. {\bf D28} (1983) 1291; 
{\bf D32} (1985) 879; 
B. Carter and R. G. Mclenaghan, {\it Generalised master equations 
for wave equation separation in a Kerr or Kerr-Newman black hole 
background}, in Proceeding of the Second Marcel Grossmann Meeting 
on General Relativity, edited by R. Ruffini, (North-Holland 
Publishing Company, 1982).

\bibitem{JC}
B. P. Jensen and P. Candelas, {\it Phys. Rev}. 
{\bf D33} (1986) 1590.

\bibitem{Leaver}
E. W. Leaver, {\it J. Math. Phys}. {\bf 27} (1986) 1238.

\bibitem{LFN}
J. W. Liu, {\it J. Math. Phys}. {\bf 32} (1992) 4026; 
B. D. B. Figneiredo, M. Novello, {\it J. Math. Phys}. 
{\bf 34} (1993) 3121.

\bibitem{WC}
S. Q. Wu and X. Cai, {\it J. Math. Phys}. {\bf 40} (1999) 4538.

\bibitem{STU2}
H. Suzuki, E. Takasugi and H. Umetsu, {\it Prog. Theor. Phys}. 
{\bf 100} (1998) 491; {\bf 102} (1999) 253.

\bibitem{Ron} 
{\sl Heun's Differential Equations}, edited by 
A. Ronveaux, (Oxford Science Publications, 1995). 

\bibitem{Heun} 
K. Heun, {\it Math. Ann}. {\bf 33} (1889) 161.

\bibitem{STM}
S. Mano, H. Suzuki and E. Takasugi, {\it Prog. Theor. Phys}. 
{\bf 95} (1996) 1079; {\bf 96} (1996) 549; 
H. Suzuki and E. Takasugi, {\it Prog. Theor. Phys}. 
{\bf 97} (1997) 213.

\bibitem{MS}
T. Masuda and H. Suzuki, {\it J. Math. Phys}. 
{\bf 38} (1997) 3669.

\bibitem{LWEV}
C. G. Lambe and D. R. Ward, {\it Quart. J. Math}. 
{\bf 5} (1934) 81; 
A. Erdelyi, {\it Quart. J. Math}. {\bf 13} (1942) 107; 
G. Valent, {\it SIAM. J. Math. Anal}. {\bf 17} (1986) 688; 
math. CA/9307204. 

\bibitem{PD}
J. F. Plebanski and M. Demianski, {\it Ann. Phys}. (NY) 
{\bf 98} (1976) 98.

\bibitem{Couch}
W. E. Couch, {\it J. Math. Phys}. {\bf 22} (1981) 1849; 
{\bf 26} (1985) 2286.

\bibitem{KGK}
D. R. Brill, P. L. Chrzanowski, C. M. Pereira, E. D. Fackerell,
J. R. Isper, {\it Phys. Rev}. {\bf D5} (1972) 1913; 
D. J. Rowan and G. Stephensen, {\it J. Phys. A: Math. Gen}. 
{\bf 10} (1977) 15; 
S. Detweiler, {\it Phys. Rev}. {\bf D22} (1980) 2323.

\bibitem{KS} 
A. Sen, {\it Phys. Rev. Lett}. {\bf 69} (1992) 1006; 
A. Garcia, D. Galtsov and O. Kechkin, {\it Phys. Rev. Lett}. 
{\bf 74} (1995) 1276.

\bibitem{Okai}
T. Okai, {\it Prog. Theor. Phys}. {\bf 92} (1994) 47.

\bibitem{OSWF}
P. M. Morse, H. Feshbach, {\sl Methods of Theoretical Physics}, 
(McGraw-Hill, New York, 1953); 
{\sl Handbook of Mathematical Functions}, edited by M. Abramowitz 
and I. A. Stegun, 9th version, (Dover, New York, 1972).

\bibitem{Hawk}
S. W. Hawking, {\it Nature}, {\bf 248} (1974) 30; 
{\it Commun. Math. Phys}. {\bf 43} (1975) 199.

\bibitem{DRS}
T. Damour, R. Ruffini, {\it Phys. Rev}. {\bf D14} (1976) 332; 
S. Sannan, {\it Gen. Rel. Grav}. {\bf 20} (1988) 239.
 
\end{thebibliography}
\end{document}